\documentclass[12pt,a4paper,reqno]{article}
\usepackage{amsmath}
\usepackage{amssymb}
\usepackage{amsthm}
\usepackage{graphicx}
\usepackage[dvips]{color}

\renewcommand{\(}{\left(}
\renewcommand{\)}{\right)}

\newcommand{\e}{\mbox{e}}
\renewcommand{\hbar}{\hslash}

\newcommand{\ox}{\otimes}
\newcommand{\<}{\langle}
\renewcommand{\>}{\rangle}

\newcommand{\be}{\begin{equation}}
\newcommand{\ee}{\end{equation}}

\theoremstyle{plain} 

\theoremstyle{definition}

\theoremstyle{remark}

\begin{document}
\title{Philosophical lessons of entanglement}
\author{Anthony Sudbery\\[10pt] \small Department of Mathematics,
    University of York,\\[-2pt] \small Heslington, York, England YO10 5DD\\ 
    \small  Email:  as2@york.ac.uk}
\date{Talk given at\\75 Years of Quantum Entanglement\\[2pt]Kolkata, India, 10 January 2011}
\maketitle

\begin{abstract}

The quantum-mechanical description of the world, including human observers, makes substantial use of entanglement. In order to understand this, we need to adopt concepts of truth, probability and time which are unfamiliar in modern scientific thought. There are two kinds of statements about the world: those made from inside the world, and those from outside. The conflict between contradictory statements which both appear to be true can be resolved by recognising that they are made in different perspectives. Probability, in an objective sense, belongs in the internal perspective, and to statements in the future tense. Such statements obey a many-valued logic, in which the truth values are identified as probabilities.

\end{abstract}

\section*{Introduction}

There is nothing new or surprising in the idea that quantum theory has philosophical implications. The theory itself, in its commonest formulation, was shaped by the philosophical predilections of Bohr and Heisenberg, and in turn was used by them to support their philosophical ideas. However, these ideas, positivist and instrumentalist in tendency, discouraged any attempt to take quantum mechanics literally, and had the effect of restricting its scope to microscopic phenomena. In Bohr's famous words, ``There is no quantum world. There is only an abstract physical description. It is wrong to think that the task of physics is to find out how nature is. Physics concerns what we can say about nature'' \cite{Bohr:quote}. But as quantum predictions are confirmed for larger and larger objects, this metaphysical restraint comes to look less and less satisfactory; it becomes ever more tempting to believe that there \emph{is} a quantum world, and that it is the whole world, including ourselves. It then becomes necessary to confront the philosophical problem ``How \emph{can} the world be as described by quantum theory?'' In particular, how can we understand entanglement as a real feature of the world? This question, I will argue, teaches us philosophical lessons about some of our basic concepts: truth, probability, time and chance.

By ``the world as described by quantum theory'' I mean a world that is associated with a single vector in Hilbert space, changing in time according to the Schr\"odinger equation, and changing \emph{only} in that way. This is (at least part of) the Everett-Wheeler interpretation of quantum mechanics, but it remains a partial description until some kind of meaning is given to the universal state vector. Notoriously, the Everett-Wheeler interpretation has problems in incorporating probability; possibly less notorious, but no less serious, are problems associated with the past and the future which were pointed out by Bell \cite{Bell:cosmologists}. Accepting the universal quantum description, with its essential feature of entanglement, requires hard thought about probability and the future. These are already mysterious enough; but I will suggest that quantum mechanics provides a framework which helps with the old mysteries. In addition, in order to make sense of entanglement in the quantum description of the whole universe, including ourselves and our thoughts and beliefs, we will need to examine what we mean by \emph{truth}.

\section*{Schr\"odinger's Cat}

We are celebrating the 75th anniversary of Schr\"odinger's paper \cite{Schrcat} in which he named and discussed the concept of entanglement, emphasising its responsibility for the strangeness of the world revealed by quantum mechanics. The concept itself, of course, is older than this; it played a central role in von Neumann's 1932 analysis of measurement \cite{vonNeumann:QM} as well as the EPR paper of 1935 to which Schr\"odinger was responding \cite{EPR}, and its essential features were already recognised in Weyl's textbook of 1928 \cite{Weyl:QM}. However, it is Schr\"odinger's paper that brings entanglement to the centre of the stage. 

That same paper contains the famous example of Schr\"odinger's cat, which is often presented as an argument against quantum mechanics (though this is not how Schr\"odinger intended it). When the unfortunate cat has been in Schr\"odinger's diabolical device for a time $t$, the crude argument goes, quantum mechanics predicts that its state is of the form
\be\label{Schrcat}
|\psi_\text{cat}(t)\> = \e^{-\gamma t}|\text{alive}\> + \sqrt{1 - \e^{-2\gamma t}}|\text{dead}\>
\ee
So why don't we see such superpositions of live and dead cats? I repeat, this is not Schr\"odinger's question: he devised this example to show that superposition could not be understood as a kind of smearing out or jellification of the individual terms in the superposition. Indeed, the answer to this question is implicit in Schr\"odinger's discussion of entanglement later in the paper. Quantum mechanics does \emph{not} predict that we will see the state \eqref{Schrcat}; if we are watching the cat, hoping to see a superposition like the above, the interaction by which we see it actually produces the entangled state
\be\label{catobserved}
|\Psi(t)\> = \e^{-\gamma t}|\text{alive}\>_\text{cat}|\stackrel{\centerdot\;\centerdot}{\smile}\>_\text{observer} + \sqrt{1 - \e^{-2\gamma t}}|\text{dead}\>_\text{cat}|\stackrel{\centerdot\;\centerdot}{\frown}\>_\text{observer}
\ee
in which $|\stackrel{\centerdot\;\centerdot}{\smile}\>$ is the observer state of seeing a live cat and $|\stackrel{\centerdot\;\centerdot}{\frown}\>$ is the state of seeing a dead cat. Nowhere in this total state is there an observer seeing a superposition of a live and a dead cat. 

But then, what \emph{does} this state tell us about the cat and the observer?

\section*{What is Truth?}

If the observer is watching the cat continuously over the period from time $0$ to time $t$, they will be able to note the time, if any, at which they see the cat die. Then the joint state of the cat and the observer is something like
\begin{multline}
|\Psi(t)\> = \e^{-\gamma t}|\text{alive}\>_\text{cat}|\text{``The cat is alive''}\>_\text{observer}\\ + \int_0^{t'}\e^{-\gamma t'}|\text{dead}\>_\text{cat}
|\text{``I saw the cat die at time $t'$''}\>_\text{observer}dt'
\end{multline}
in which the observer states contain propositions which are physically encoded in the brain of the observer. But what is their status as propositions; are they true or false?  Each is believed by a brain which has observed the fact it describes, and that fact belongs to reality. As a
human belief, each statement could not be more true. Yet they cannot all be true, for they contradict each other.

This conflict shows the necessity of considering the context in which a statement is made when discussing its truth value. When this is done, it becomes possible for contradictory statements to be simultaneously true, each in its own context. 

\section*{Internal vs. External}

In general, the state of the universe can be expanded in terms of the states of any observer inside the universe as
\[
|\Psi(t)\> = \sum_n|\eta_n\>|\Phi_n(t)\>
\]
where the $|\eta_n\>$ form an orthonormal basis of observer states, which we can take to be eigenstates of definite experience; the $|\Phi_n(t)\>$ are the corresponding states of the rest of the universe at time $t$. The actual observer can only experience being in one of the states $|\eta_n\>$ (because they exhaust all possible experiences), and in this state it is true for the observer that the only experience they have is $\eta_n$; the observer is justified, at time $t$, in deducing that the rest of the universe is in the unique state $|\Phi_n(t)\>$. This is the \emph{internal} truth relative to the experience state $|\eta_n\>$.

But there is also the \emph{external} truth that the state of the whole universe is $|\Psi(t)\>$. From this standpoint all the experiences $\eta_n$ truly occur. 
Thus there are the following two types of truth involved.

{\bf External truth:} The truth about the universe is given by a state vector $|\Psi(t)\>$ in a Hilbert space $\mathcal{H}_U$, evolving according to the Schr\"odinger equation. If the Hilbert space can be factorised as 
\[
\mathcal{H}_U = \mathcal{H}_S\ox\mathcal{H}_E
\]
where $\mathcal{H}_S$ contains states of an experiencing observer, then 
\[
|\Psi(t)\> = \sum_n|\eta_n\>|\Phi_n(t)\>
\]
and all the states $|\eta_n\>$ for which $|\Phi_n(t)\> \ne 0$ describe experiences which actually occur at time $t$.

\bigskip

{\bf Internal truth} from the perspective $|\eta_n\>$: I actually have experience $\eta_n$, which tells me that the rest of the universe is in the state $|\Phi_n(t)\>$. This is an objective fact; everybody I have talked to agrees with me.

\section*{Compatibilism}

This distinction between internal and external truth is a special case of a general opposition which was introduced and carefully discussed by the philosopher Thomas Nagel \cite{Nagel:nowhere}. He used it to discuss a number of longstanding conflicts in philosophy, including
\begin{enumerate}
\item The existence of space-time \emph{vs} the passage of time;
\item Determinism \emph{vs} free will;
\item The physical description of brain states \emph{vs} conscious experience;
\item Duty \emph{vs} ``Why should I?"
\end{enumerate}
These are all contradictions between pairs of statements or principles, both of which we seem to have good reason to believe. In every case one of the statements is a general universal statement --- what Nagel \cite{Nagel:nowhere} calls ``a view from nowhere" --- to which assent seems to be compelled by scientific investigation or moral reflection; the other is a matter of immediate experience, seen from inside the universe (a view from ``now here").

 Scientists might be tempted to exalt the external statement as the objective truth, downgrading internal statements as merely subjective. Indeed, Nagel himself uses the terminology of ``objective'' and ``subjective''. But he does not use a dismissive qualifier like 
 ``merely'' to denigrate the subjective: he is at pains to emphasise that the truth of an internal statement has a vividness and immediacy, resulting from the fact that it is actually experienced, compared to which external truth is ``bleached-out''. This applies most obviously in contexts like ethics and aesthetics, but we would do well to remember it in our scientific context; as I have pointed out above, it is the internal statement which has the scientific justification of being supported by evidence, and is objective in the usual sense that it is empirical and is agreed by all observers who can communicate with each other.

But the situation is more complicated than this might suggest. It is not that there is a God-like being who can survey the whole universe and make statements about the universal state vector, distinct from us physical beings who are trapped in one component of $|\Psi\>$. It is we physical beings who make statements about $|\Psi\>$, for good theoretical reasons, from our situation in which we experience just the one component $|\eta_n\>|\Phi_n\>$. From that perspective, what are we to make of the other components $|\eta_m\>|\Phi_m\>$?

\section*{``$+$'' $=$ ``and'' or ``or'' or `` `and' and `or' ''?}

Consider a measurement process, in which an initial state $|\Psi(0)\> = $ $|\eta_0\>|\Phi_0(0)\>$, containing only one experience $|\eta_0\>$, develops in time $t$ to an entangled state $|\Psi(t)\> = \sum|\eta_n\>|\Phi_n(t)\>$. The external statement is: 
\begin{quotation} $|\Psi(t)\>$ represents a \emph{true} statement about the universe, and all its components are \emph{real}. \end{quotation}
The observer who experiences only $|\eta_n\>$ must say: 
\begin{quotation} I know that only $|\eta_n\>$ is {\bf real} (because I experience only that), and therefore $|\Phi_n(t)\>$ represents a {\bf true} statement about the rest of the universe. But I also know that $|\Psi(t)\>$ is \emph{true} (because I've calculated it). The other $|\eta_m\>|\Phi_m\>$ represent things that {\bf might have happened} but {\bf didn't}.
\end{quotation}
These statements are font-coded, using bold type for internal (vivid, experienced) judgements, and italic for external (pale, theoretical) ones, even though these are made by an internal observer.

It is a constant temptation in physics, on finding a quantum system in a superposition $|\phi\> + |\psi\>$, to think that it is either in the state $|\phi\>$ or in the state $|\psi\>$. This, after all, is the upshot when we look at the result of an experiment. Despite the stern warnings of our lecturers when we are learning the subject, and the proof from the two-slit experiment that ``$+$'' cannot mean ``or'', we all probably slip into this way of thinking at times; and the common-sense view of Schr\"odinger's cat seems to justify it. On the other hand, the many-worlds view insists that both terms in the superposition are real, and therefore ``$+$'' means ``and''. What I am suggesting here is that both ``and'' and ``or'' are valid interpretations of ``$+$'' in different contexts: ``and'' in the external view, ``or'' in the internal view. 

\section*{How Many Worlds?} 

The foregoing needs some refinement. Quantum superposition is not just a single binary operation ``$+$'' on states, but is modified by coefficients: $a|\phi\> + b|\psi\>$ is a weighted superposition of the (normalised) states $|\phi\>$ and $|\psi\>$. In interpreting ``$+$'' as ``or'', it is easy to incorporate this weighting of the disjuncts by interpreting it in terms of probability. But if we interpret ``$+$'' as ``and'', as in the many-worlds interpretation, what can it mean to weight the conjuncts? Let us look back to the paradigmatic geometrical meaning of vector addition. ``Going north-east'' is the vector sum of ``going north'' and ``going east'', and does indeed mean going north \emph{and} going east at the same time. But ``going NNE'' also means going north and going east at the same time; only there is more going north than going east. So if a superposition $a|\phi\> + b|\psi\>$ of macroscopic states $|\phi\>$ and $|\psi\>$ means that both $|\phi\>$ and $\psi\>$ are real in different worlds, we must accept that they are not ``equally real'', as is often carelessly stated (see, for example, the blurb of \cite{Byrne}), but that they are real to different extents $|a|^2$ and $|b|^2$. Adding up these degrees of reality, we then find that there are not many worlds but ($|a|^2 + |b|^2 =$) one.

Another argument for this conclusion uses the physical observable of particle number. The many-worlds view regards the state vector $|\Psi(t)\> = \sum_n|\eta_n\>|\Phi_n(t)\>$ as describing many (say $N$) worlds, with $N$ different copies of the observer having the different experiences $|\eta_n\>$. Suppose the observer's name is Alice. There is an observable called Alice number, of which each of the states $|\eta_n\>|\Phi(t)\>$ is an eigenstate with eigenvalue 1. Then $\Psi(t)\>$ is also an eigenstate of Alice number with eigenvalue 1 (not $N$). There is only one Alice.

\section*{Collapse}

The observer in this measurement process might go on to say:
\begin{quotation}
I \textbf{saw} a transition from $|\eta_1\>$ to $|\eta_n\>$ at some time $t' < t$. But I \emph{know} that $|\Psi(t')\>$ didn't collapse. The other $|\eta_m\>|\Phi_m\>$ \textbf{might} come back and interfere with me in the future; but this has very low \textbf{probability}.
\end{quotation}
This shows how the conflicting statements about time development in the Everett-Wheeler and Copenhagen interpretations can after all be compatible. The continuous Schr\"odinger-equation evolution postulated by the Everett-Wheeler interpretation refers to the external view; the collapses postulated by the Copenhagen interpretation refer to the internal view, i.e.\ to what we actually see. The occurrence of collapse becomes a theorem rather than a postulate (conjecturally --- more on this later). We also see that it is only the internal statement that mentions probability. But what does it mean?

'\section*{What is Probability?}

The meaning of probability is a long-standing philosophical problem (see, for example, \cite{Gillies}). It seems likely that there are in fact several distinct concepts which go by the name of probability, sharing only the fact that they obey the same mathematical axioms. The clearest of these, perhaps, is degree of belief, which has the advantage that it can be defined operationally: someone's degree of belief in a proposition is equal to the odds that they are prepared to offer in a bet that the proposition is true. The subjective nature of this concept seems to chime with the fact that it belongs in internal statements, as we have just seen, and indeed similar views of probability are often adopted by Everettians even though their general stance is objectivist.

However, we have also seen that ``internal'' should not be equated with ``subjective'', and our experience in a quantum-mechanical world seems to require a description in terms of objective chance. Things happen randomly, but with definite probabilities that cannot be reduced to our beliefs. The value of the half-life of uranium 238 is a fact about the world, not a mere consequence of someone's belief. 

Such objective probability can only refer to future events. 

\section*{Probability and the Future}

What kinds of statements can be made at time $t$ about some future time $s > t$, if the universal state vector is known to be $|\Psi(t)\>$ and its decomposition with respect to experience states of a particular observer is $\sum_n|\eta_n\>|\Phi_n(t)\>$? From the external perspective, the future state $|\Psi(s)\>$ is determined by the Schr\"odinger equation and there is no question of any probability. From the internal perspective relative to an experience state $|\eta_n\>$, there is a range of possible future states $|\eta_m\>$, and probabilities must enter into the statement of what the future state will be. But here is a fundamental problem: there is \emph{no such thing} as what the future state will be. As Bell pointed out, quantum mechanics gives no connection between a component of $|\Psi\>$ at one time and any component at another time; so what is it that we can assign probabilities to? How can ``the probability that my state will be $|\eta_m\>$ tomorrow" mean anything when ``my state will be $|\eta_m\>$ tomorrow" has no meaning?

\section*{The Classical Future}

This puzzle takes us back to ways of thinking that are much older than quantum mechanics, indeed older than all of modern science. The success of Newtonian deterministic physics has led us to assume that there always is a definite future, and even when we drop determinism we tend to continue in the same assumption. There is a future, even if we do not and cannot know what it will be. But this was not what Aristotle believed, and maybe it is not what we believed when we were children. 

Aristotle, in a famous passage \cite{seabattle}, considered the proposition ``There will be a sea-battle tomorrow''.
He argued that this proposition is neither true nor false (otherwise we are forced into fatalism). Thus he rejected the law of excluded middle for future-tense statements, implying that they obey a many-valued logic. Modern logicians \cite{Prior} have considered the possibility of a third truth-value in addition to ``true" or ``false", namely $u$ for ``undetermined", for future-tense statements. But, interestingly, Aristotle admitted that the sea-battle might be more or less likely to take place. This suggests that the additional truth values needed for future-tense statements are not limited to one, $u$, but can be any real number between 0 and 1 and should be identified with the probability that the statement will come true. Turning this round gives us an objective form of probability which applies to future events, or to propositions in the future tense; in a slogan,
\[
\text{Probability} = \text{degree of truth}.
\]

\section*{Probabilities as Truth Values}

This translation of Aristotle's position seems so natural that it has surely been developed already. However, I have been unable to find it in the literature of probability theory, temporal logic or many-valued logic. The notion of ``degree of truth" occurs in fuzzy logic and philosophical discussions of vagueness (for a critical account see \cite{vagueness}), but seems to have been little used in the philosophy of probability. In an early paper \cite{Luk:prob} (earlier than what are generally regarded as the first papers on many-valued logic) \L ukasiewicz introduced truth values between 0 and 1 and equated them with probabilities, but in a different sense from that of quantum mechanics. He was searching for a notion of objective probability, but found it only in propositions containing a free variable; this is the concept of probability used by number theorists, for example, who might consider the probability that a number in an arithmetic progression is prime. \L ukasiewicz rejected any application to future-tense statements with no free variables, to which he thought probability did not apply because at this time (1913) he believed in determinism. Reichenbach \cite{Reichenbach:probability} interpreted probability as a truth value, though because he held a frequentist view of probability, his truth values were properties of sequences of propositions rather than single propositions. \L ukasiewicz defined the $[0,1]$-valued truth value of a proposition with a variable in terms of the (two-valued) truth or falsity of the singular propositions obtained by substituting individuals for the variable; Reichenbach defined the $[0,1]$-valued truth value of a sequence of propositions in terms of the two-valued truth or falsity of the propositions in the sequence. Here it is proposed that the $[0,1]$-valued truth value of a future-tense proposition is a primitive property of that proposition, not reducible to any truth values in two-valued logic.

Truth values are also equated with probabilities in the topos theory approach to quantum mechanics developed by Isham and Doering \cite{toposreview}; in their work the concept of probability is stretched to fit a concept of truth value already existing in many-valued logic (in particular, it is no longer a number in the interval $[0,1]$). Here, on the other hand, it is proposed to adapt the concept of truth value to fit the pre-existing concept of probability. Two aspects of this adaptation particularly need to be pointed out. 

First, truth values are usually assumed to have the property that logical connectives like $\land$ (and) and $\lor$ (or) are ``truth-functional'', i.e.\ the truth values of $p\land q$ and $p\lor q$ are determined by those of $p$ and $q$. The probability of a compound statement, however, is constrained but not determined by the probabilities of its constituents: for the probabilities of $p\land q$ and $p\lor q$ we only have inequalities
\begin{align*}
P(p), P(q) &\ge P(p\land q) \ge 1 - P(p) - P(q),\\
P(p), P(q) &\le P(p\lor q) \le P(p) + P(q).
\end{align*}
Although degrees of truth for vague statements are usually assumed to be truth-functional, some authors have argued that they should behave like probabilities, as above (\cite{Edgington}; see also \cite{vagueness}). Reichenbach proposed to solve this problem by increasing the number of arguments in the truth tables. The truth value of a compound proposition formed from $p$ and $q$ is a function, not only of the truth values of $p$ and $q$, but also of a third value, corresponding to the conditional probability of $p$ given $q$. This amounts to recognising the truth value of $p\land q$ as an independent variable.

A second departure from the usual properties of truth values is that the probability of a proposition referring to a future time $s$ depends not only on the proposition itself (taking the time $s$ to be a part of the proposition), but also on the time $t$ at which the proposition is considered. For an observer experiencing the state $|\eta_n\>$ at time $t$, the probability of experiencing $\eta_m\>$ at a future time $s$ is 
\be\label{prob}
\frac{\left|\(\<\eta_m|\<\Phi_m(s)|\)\e^{-iH(s-t)/\hbar}\(|\eta_n\>|\Phi_n(t)\>\)\right|^2}{\<\Phi_m(s)|\Phi_m(s)\>\<\Phi_n(t)|\Phi_n(t)\>}.
\ee
Thus our account of probability requires that the truth value of the proposition ``My experience at time $s$ will be $|\eta_m\>$'' depends on the time at which the proposition is uttered. Such dependence on a context of utterance is nothing new (consider the truth value of the proposition ``it is raining''), but that can usually be understood as being because the context of utterance is needed to fill in an incomplete proposition (in our example, it provides the time and place at which it is raining). There is no such incompleteness in the case of the above proposition. We therefore require a framework of time-dependent truth values, which is recognised in the Stanford Encyclopaedia of Philosophy as ``the tensed view of semantics'' \cite{stanford:time}. 

\section*{The Truth of the Past}

The classical position (in classical philosophy, if not classical physics) would be that propositions referring to the past and present are either true or false; it is only the future that is uncertain. Thus if $P(s, t)$ is the truth value at $t$ of a proposition referring to time $s$, we should have
\begin{align}\label{past}
0 \le P(s, t) \le 1 &\quad \text{ if } s > t,\notag\\
P(s,t) = 0 \text{ or } 1 &\quad \text{ if } s \le t.
\end{align}
It is not clear whether this can or should be maintained in quantum theory. Bell's point \cite{Bell:cosmologists} about the lack of connection between experience states at different times applies to the past as much as the future. We do have an empirical warrant for our past states, as we do not for future states, in the form of memory, which is not symmetric under time reversal. This has yet to be modelled quantum-mechanically, but hopefully it can be shown that there is a physical process which leaves a record in the state at one time of a sequence of states in the past, and that this record is consistent with the probabilities \eqref{prob}. Maybe it would go further and establish transition probabilities such as have been postulated for modal interpretations \cite{BacciaDickson, verdammt}. If so, this might provide justification for truth values satisfying \eqref{past}. 

Even if this could be established as a feature of quantum systems with memory, the theory would still be vulnerable to Bell's charge of temporal solipsism: a memory state is still a present state, and does not constitute a genuine past. Markosian \cite{openpast} has pointed out that this openness of the past is an inevitable consequence of time-reversal symmetry in a theory with an open future. 

\section*{The Open Future}

We find it hard, in a scientific theory, to accommodate the idea that there is no definite future. To be sure, we have indeterministic theories in which the future is not uniquely determined by the past, but such stochastic theories deal with complete histories encompassing past, present and future; probabilities refer to which of these histories is actual. Indeterminism, in the usual stochastic formulation, consists of the fact that there are many such histories containing a given past up to a certain time, so the future extension is not unique; but the underlying assumption is that only one of these future histories is real, so that the future is fixed even though it is not determined. In contrast, the formulation of quantum mechanics outlined here --- or what Bell \cite{Bell:cosmologists} called the ``Everett (?) theory'' --- is, I think, the only form of scientific theory in which the future is genuinely open. Unlike Bell, I do not regard this as a problem for the theory; it tells a truth which we should be glad to recognise. The function of the theory is to provide a catalogue of possibilities and specify how these change (deterministically) with time; it does not and cannot say which of the possibilities is actualised at any time. The ``measurement problem'' of quantum theory is no more than the difficulty of accepting this format for a scientific theory; with a change of gestalt, we can see it as a natural way to formulate indeterminism.

However, I must emphasise the roles that entanglement and the concept of internal truth play in this resolution of the measurement problem. Without these, there would be a ``preferred basis'' problem: if the universal wave function is a catalogue of possibilities, what basis defines the components which are to be regarded as possibilities? But there is no preferred-basis problem in this understanding of the Everett-Wheeler interpretation. The possibilities are given by experience states, which only exist if the universal Hilbert state has a tensor product structure in which one of the factors describes a system capable of experience, i.e. which has a basis of states exhibiting the structure of propositions describing experience. It is not required that this basis should be unique; it is in principle possible that the same factor Hilbert space might have a different basis also showing the structure of a (totally different) set of experiences. It is also in principle possible that the universal Hilbert space has more than one tensor product structure with the required properties. If this should be so, statements about these different experiences would also be (internally) true, relative to these different structures; this would not detract from the truth of the original experience propositions. Both kinds of internal proposition would be compatible with the external truth of the same universal state vector. 


\section*{Summary: The lessons of entanglement}

I conclude with a set of slogans.

\begin{itemize}
\item There are different kinds of truth. We must distinguish external and internal truth.
\item The logic of future-tense statements is many-valued.
\item Probability = degree of truth.
\item Entanglement captures indeterminism.
\end{itemize}

\bibliographystyle{aipprocl}
\bibliography{quantum}

%
%


\end{document}